# EXTENDING THE REACH OF QKD USING RELAYS

*S.M. Barnett*[1] *& S.J.D. Phoenix*[2]

[1]Dept of Physics, Strathclyde University, Glasgow, UK
[2]Khalifa University, Abu Dhabi Campus, PO Box 127788, Abu Dhabi, UAE

**ABSTRACT**

One of the obstacles to deployment of QKD solutions has been the distance limitation. Solutions using relays have been proposed but these rely on link-by-link key establishment. We present a new technique to extend the distance of a quantum key distribution channel using an active relay. Each relay acts as an intercept/resend device and allows the establishment of an end-to-end key. It has been argued that such relays cannot be used to extend the distance, but we show that with a suitable adaptation of the protocol the effective key distribution distance can be increased.

## 1. INTRODUCTION

From its beginnings as a theoretical curiosity some two decades ago Quantum Key Distribution (QKD) is now a commercially available technology that is currently undergoing trials in a number of locations [1]. The technology offers a way of establishing a random sequence of binary digits between two end users in such a way that the secrecy of the established bit strings can be guaranteed. This bit string can then be used as a key in cryptographic applications.

One of the biggest obstacles to the widespread introduction of QKD techniques is the distance limitation in optical fibre which restricts current applications to a few tens of kilometers. The distance can be extended by using relays. In standard commercial applications these relays establish link-by link keys. As Bechmann-Pasquinucci and Pasquinucci have shown [2] it is possible to use relays in an end-to-end mode in which the relays simply act as intercept/resend devices passing on the state they measure. However, it has been argued in [2] that this cannot increase the distance for a QKD transmission over that of a standard channel without relays. The key transmission rate also decreases exponentially with the number of relays.

We show that neither of these limitations for end-to-end relays persists with a modification of the protocol. Furthermore we show that the fraction of transmitted data that can be used for key establishment is independent of the number of relays and is precisely equal to ½, that is, the same fraction as for a standard QKD channel without relays.

This result shows that end-to-end keys can be established over significant distances using these relays. Elsewhere [3] we have shown how these relays can be secured using a secret sharing technique. Taken together these methods describe a practical and secure implementation of QKD for the transmission of quantum keys over long distances.

## 2. LIMITATIONS OF INTERCEPT/RESEND RELAYS FOR QKD

We consider a channel over which Alice and Bob wish to exchange keys using QKD. The distance between them, however, is too great to do this using a normal transmission and so one or more relays $R_n$ must be used

The intercept/resend technique described in [2] allows the establishment of end-to-end keys. The protocol described there, however, cannot be used to extend the distance of the QKD transmission. In effect the authors consider the entire communication Alice – Relay – Bob as a single communication. If the distance between Alice and the relay is only just sufficient for a successful quantum key exchange the photon loss for the second half of this channel is clearly too great to establish a successful QKD transmission using this protocol.

The relay, or node $R_1$ using the above terminology, simply measures the state of the incoming photons as in the usual QKD protocol and re-transmits the photons in the measured state. This process is repeated for all of the intermediate nodes until the destination point on the channel.

Let us consider a simple channel consisting of Alice, relay node $R_1$ and Bob. In the BB84 protocol [4] there are two alphabets which are formed from the eigenstates of complementary operators. Let us call these two operators $\hat{X}$ and $\hat{Y}$ with eigenstates $|\pm\rangle_X$ and $|\pm\rangle_Y$ respectively. There are 8 possible measurement chains for each photon. For example, Alice could choose to encode using the $\hat{X}$ basis alphabet, relay node $R_1$ could choose to measure and retransmit in the $\hat{Y}$ basis alphabet and Bob could choose to measure in the $\hat{Y}$ basis alphabet. The possible variations are shown in Table 1.

**Table 1.** The possible measurements and codings for a simple 3 node channel. The 'key' column indicates that the measurements can lead to establishment of a key between the nodes listed. Thus for the instances described in the first two rows where all three nodes have used the same alphabet or coding scheme then all parties can use these photons to establish a key. In row 3 and 4 it is seemingly only possible to use these measurements to establish a key between Alice and node $R_1$. It is not possible to establish a quantum key using the measurements indicated in the last two rows. Note that in this case only ¼ of the transmissions are usable by Alice and Bob to establish a key between them, as opposed to ½ in the standard protocol without an intermediary. However, this is offset by the fact that overall ¾ of the transmissions are useful with the addition of the intermediary as opposed to only ½ in the usual protocol. This is a critical factor in the adaptation of the protocol to enhance the key rate that we outline in section 4.

| Alice | Relay $R_1$ | Bob | Key |
|-------|-------------|-----|-----|
| X | X | X | A - $R_1$ -B |
| Y | Y | Y | A - $R_1$ -B |
| X | X | Y | A – $R_1$ |
| Y | Y | X | A – $R_1$ |
| X | Y | Y | $R_1$ – B |
| Y | X | X | $R_1$ – B |
| X | Y | X | *No key* |
| Y | X | Y | *No key* |

By extending this to a chain of $n$ nodes, that is the two endpoints Alice and Bob and $n - 2$ trusted intermediate relay nodes we see that the only transmissions where no useful data can be obtained are those which have an alternating pattern of measurements so that the fraction of useful transmissions is given by

$$f_n = 1 - \frac{1}{2^{n-1}} \quad (1)$$

In all other cases a key can be established somewhere between two nodes. We see from equation (1) that as $n$ increases the useful fraction approaches 1. However, as $n$ increases the number of transmissions that are useful to the two endpoints Alice and Bob decreases. The only transmissions where Alice and Bob can establish a key are those in which all nodes have chosen the same basis and thus the fraction of transmissions that can be used to establish an end-to-end key is $1 - f_n$.

It would seem from the above discussion that there are two apparent limitations on using relays in this 'pass-through' mode. Firstly, it appears that they cannot be used to extend the distance for a QKD channel and secondly, it appears that the fraction of timeslots that can be used to establish an end-to-end key decreases exponentially with the number of intermediate relays.

The first of these limitations arises because the signal to noise performance as the distance extends ensures that the quantum signal cannot be distinguished from noise. This would certainly be the case if all the relays did was to operate according to the protocol outlined. However, as we now show, if the relay protocol is adapted the argument outlined in [2] to demonstrate the impossibility of distance extension for QKD using these pass-through relays is invalid.

## 3. OVERCOMING THE DISTANCE LIMITATION

Let us consider a channel in which there is only one relay $R_1$. Let us further suppose that the distance between Alice and Bob is too great for them to establish a quantum key without some kind of assistance. There will be some loss factor $\xi$ on the channel between Alice and $R_1$ that means that the relay no longer has sufficient data points to resend to overcome the SNR problems for the second hop on the channel.

When we view the channel as two separate hops like this the solution becomes obvious. The relay must operate in such a way to increase the SNR performance for the second hop. There are two ways to achieve this. The relay can simply wait to resend until sufficient data points have been collected to allow the resend transmission to proceed at an acceptable data rate to overcome the SNR limit for the second hop.

The second solution is for the relay to use a padding technique. In the timeslots where nothing is received by $R_1$ a new photon is transmitted encoded in a random basis with a random bit value as in the standard two-user QKD channel. The relay just keeps a record of the timeslots that are padded and those that are resends from the data received from Alice. At the end of the channel Bob now receives a signal of sufficient strength to overcome the SNR limitations but only a fraction of these, governed by $\xi$, have originated from Alice. One advantage of the padding technique over the delay method is that it allows the establishment of a pool of key material between $R_1$ and Bob which may be useful.

As the number of relays, and therefore the number of hops, is increased we see that the fraction of data received by Bob that originates with Alice scales as $\xi^{-m}$ where $m$ is the number of hops. This will ultimately limit the practical operational distance for QKD relays by slowing the Alice – Bob key rate to an unacceptably low level. We shall discuss the practical operational parameters for these relay channels elsewhere.

## 4. OVERCOMING THE EXPONENTIAL REDUCTION OF THE USEFUL DATA

As we have seen, the introduction of a pass-through relay operating as a simple intercept/resend device leads to an apparent reduction in the eventual key rate between Alice and Bob, as described by equation (1). In order to illustrate that this is not, in fact, the case we consider a channel in which there are two relays, $R_1$ and $R_2$. In Table 2 and Table 3 we show the possible bit values for the timeslots when the participants in the channel use the coding bases $\hat{X} - \hat{X} - \hat{Y} - \hat{Y}$ and $\hat{X} - \hat{Y} - \hat{Y} - \hat{X}$, respectively.

It is important to note that the relays, unlike an eavesdropper, are not adversaries, but can act as cooperative entities assisting the eventual establishment of a secret key between Alice and Bob. It is this critical detail that allows the relays to cooperate in such a way as to overcome the apparent problem of the overall key reduction factor given in equation (1).

Table 2. The possible bit values transmitted when the channel participants adopt the coding scheme, $\hat{X} - \hat{X} - \hat{Y} - \hat{Y}$. Each row corresponds to a possible timeslot for the QKD transmission.

| Alice ($\hat{X}$) | $R_1$ ($\hat{X}$) | $R_2$ ($\hat{Y}$) | Bob ($\hat{Y}$) |
|---|---|---|---|
| 1 | 1 | 1 | 1 |
| 1 | 1 | 0 | 0 |
| 0 | 0 | 1 | 1 |
| 0 | 0 | 0 | 0 |

Table 3 The possible bit values transmitted when the channel participants adopt the coding scheme, $\hat{X} - \hat{Y} - \hat{Y} - \hat{X}$. Each row corresponds to a possible timeslot for the QKD transmission.

| Alice ($\hat{X}$) | $R_1$ ($\hat{Y}$) | $R_2$ ($\hat{Y}$) | Bob ($\hat{X}$) |
|---|---|---|---|
| 1 | 1 | 1 | 1 |
| 1 | 1 | 1 | 0 |
| 1 | 0 | 0 | 1 |
| 1 | 0 | 0 | 0 |
| 0 | 1 | 1 | 1 |
| 0 | 1 | 1 | 0 |
| 0 | 0 | 0 | 1 |
| 0 | 0 | 0 | 0 |

Recall that what we are trying to achieve with QKD is the establishment of a shared sequence of secret random bits between Alice and Bob. So all we need to do is to ensure that the correct bit value is established at each hop of the channel. Accordingly, we can pair these timeslots, as desired, in order to establish a continuous bit path through the channel. This process is done post-transmission when the data points have been collected.

In the 2$^{nd}$ and 3$^{rd}$ columns of Table 2 we see that the two relays have chosen different bases. In normal QKD protocols this would be rejected as incapable of transmitting the key. However, we see in Table 3 that this gap is 'bridged' because the relays have chosen the same coding basis. So in order to establish the bit value from Table 2 at the second relay the first relay now chooses a suitable timeslot from Table 3 and informs the second relay that this timeslot is to be used. The original bit value transmitted by Alice can therefore be established. In effect this is equivalent to a bit flipping by the relays at suitable points where the bit here determines which coding basis to use.

Providing we can draw a continuous path of legitimate QKD transmission through the communication we can use this bit flipping to propagate the key through the channel. As in the standard QKD system the actual bit value coded on the photon is never revealed, only the basis and timeslot information. The key is thus propagated in secret.

It is clear upon reflection that each timeslot, except those originating from $\hat{X} - \hat{Y} - \hat{X} - \hat{Y}$ or $\hat{Y} - \hat{X} - \hat{Y} - \hat{X}$, can be paired in this fashion in the post-processing. The key reduction factor is therefore precisely the same as that for a single QKD channel with only two users. In other words, the introduction of relays does not affect the amount of potential key data.

## 5. CONCLUSION

We have shown how the distance limitation of QKD can be overcome by the use of relays operated in an intercept-resend mode in order to establish a true end-to-end key between Alice and Bob. It has also been established that the introduction of such relays does not affect the amount of data that can be used to form the eventual key. This is an important new technique that will allow the establishment of a QKD system over long distances.